\documentclass[11pt,a4paper]{article}
\usepackage{amsmath,amsfonts,graphicx,natbib,psfrag,color}
\usepackage{algorithm}

\newcommand{\Hag}{\textrm{\sf Hag}}
\newcommand{\hag}{\textrm{\sf hag}}
\newcommand{\codY}{\textrm{\sf codY}}
\newcommand{\CodY}{\textrm{\sf CodY}}
\newcommand{\SigD}{\textrm{\sf SigD}}
\newcommand{\flache}{\textrm{\sf flache}}
\newcommand{\SigDhag}{\textrm{\sf SigD\_hag}}
\newcommand{\CodYhag}{\textrm{\sf CodY\_hag}}
\newcommand{\CodYflache}{\textrm{\sf CodY\_flache}}

\title{Bayesian inference for Markov jump processes with informative observations}
\author{Andrew Golightly\thanks{email: \texttt{andrew.golightly@ncl.ac.uk}} \and\ Darren J.~Wilkinson\thanks{email: \texttt{darren.wilkinson@ncl.ac.uk}}}
\date{School of Mathematics \& Statistics, Newcastle University,\\
  Newcastle-upon-Tyne, NE1 7RU, UK}

\begin{document}
\maketitle
\begin{abstract}
In this paper we consider the problem of parameter inference for Markov jump 
process (MJP) representations of stochastic kinetic models. Since transition 
probabilities are intractable for most processes of interest yet forward simulation 
is straightforward, Bayesian inference typically proceeds through computationally 
intensive methods such as (particle) MCMC. Such methods ostensibly require the 
ability to simulate trajectories from the conditioned jump process. When 
observations are highly informative, use of the forward simulator is likely 
to be inefficient and may even preclude an exact (simulation based) analysis. 
We therefore propose three methods for improving the efficiency of simulating 
conditioned jump processes. A conditioned hazard is derived based on an 
approximation to the jump process, and used to generate end-point conditioned 
trajectories for use inside an importance sampling algorithm. We also adapt a recently proposed 
sequential Monte Carlo scheme to our problem. Essentially, trajectories are 
reweighted at a set of intermediate time points, with more weight assigned to 
trajectories that are consistent with the next observation. We consider two 
implementations of this approach, based on two continuous approximations of 
the MJP. We compare these constructs for a simple tractable jump process before 
using them to perform inference for a Lotka-Volterra system. The best performing 
construct is used to infer the parameters governing a simple model of 
motility regulation in \emph{Bacillus subtilis}. 
\end{abstract}

\noindent\textbf{Keywords:} Bayes; chemical Langevin equation (CLE); linear noise approximation (LNA); 
Markov jump process (MJP); pMCMC; particle marginal Metropolis-Hastings (PMMH); 
sequential Monte Carlo (SMC); stochastic kinetic model (SKM).

\section{Introduction}
\label{sec:intro}

Stochastic kinetic models, most naturally represented by Markov jump processes (MJPs), 
can be used to model a wide range of real-world phenomena including the 
evolution of biological systems such as intra-cellular processes 
\citep{GoliWilk05,Wilkinson09}, predator-prey interaction \citep{BWK08,ferm2008,GoliWilk11} 
and epidemics \citep{bailey1975,oneill1999,boys2007}. The focus of this paper 
is to perform exact and fully Bayesian inference for the parameters governing the MJP, 
using discrete time course observations that may be incomplete and subject to measurement error. 
A number of recent attempts to address the inference problem have been made. For 
example, a data augmentation approach was adopted by \cite{BWK08} and applied to 
discrete (and error-free) observations of a Lotka-Volterra process. The particle 
marginal Metropolis-Hastings (PMMH) algorithm of \cite{andrieu09} has been applied by 
\cite{GoliWilk11} and \cite{sherlock2014} to estimate the parameters in model auto-regulatory 
networks. 

The PMMH algorithm offers a promising approach, as it permits a joint update 
of the parameters and latent process, thus alleviating mixing problems associated with 
strong correlations. Moreover, the simplest approach is ``likelihood-free'' in the sense that 
only forward simulations from the MJP are required. These simulations can be readily 
obtained by using, for example, Gillespie's direct method \citep{Gillespie77}. The 
PMMH scheme requires running a sequential Monte Carlo (SMC) scheme (such as the 
bootstrap particle filter of \cite{gordon93}) at every iteration. Given the potential 
for huge computational burden, improvements to the overall efficiency of PMMH for MJPs has 
been the focus of \cite{Goli14}. The latter propose a delayed acceptance 
analogue of PMMH, (daPMMH), that uses approximations to the MJP such as the chemical Langevin 
equation (CLE) and linear noise approximation (LNA) \citep{kampen2001} to screen out 
parameter draws that are likely to be rejected by the sampler. It should be noted that 
the simplest likelihood free implementations of both PMMH and daPMMH are likely to 
perform poorly unless the noise in the measurement error process dominates the 
intrinsic stochasticity in the MJP. Essentially, in low measurement error cases, 
only a small number of simulated trajectories will be given reasonable weight 
inside the SMC scheme, leading to highly variable estimates of marginal likelihood 
used by the PMMH scheme to construct the acceptance probability. Intolerably 
long mixing times ensue, unless computational budget permits a large number of 
particles to be used. In the special case of error-free observation, the algorithm 
will be impracticable for models of realistic size and complexity, since in this case, 
trajectories must ``hit'' the observations.

The development of efficient schemes for generating MJP trajectories that are conditioned 
on the observations, henceforth referred to as MJP bridges, is therefore of 
paramount importance. Whilst there is considerable work on the construction 
of bridges for continuous valued Markov (diffusion) processes 
\citep{DurhGall02,DelHu06,Fearnhead08,StramYan07,schauer14,delmoral14}, seemingly little has been done 
for discrete state spaces. The approach taken by \cite{BWK08} linearly 
interpolates the hazard function between observation times but requires full 
and error-free observation of the system of interest. \cite{FanShe08} consider 
an importance sampling algorithm for finite state Markov processes where informative 
observations are dealt with by sampling reaction times from a truncated exponential 
distribution and reaction type probabilities are weighted by the probability of 
reaching the next observation. \cite{hajiaghayi14} improve the performance of 
particle-based Monte Carlo algorithms by analytically marginalising waiting times. 
The method requires a user-specified potential to push trajectories towards 
the observation.            

Our novel contribution is an MJP bridge obtained by sampling a jump process 
with a conditioned hazard that is derived by approximating the expected 
number of reaction events between observations, given the observations themselves. 
The resulting hazard is time dependent, however, we find that a simple implementation 
based on exponential waiting times between proposed reaction events works well 
in practice. We also adapt the recently proposed bridge particle filter of 
\cite{delmoral14} to our problem. Their scheme works by generating forward 
simulations from the process of interest, and reweighting at a set of 
intermediate times at which resampling may also take place. A look ahead step 
in the spirit of \cite{lin2013} prunes out trajectories that are inconsistent 
with the next observation. The implementation requires an approximation to 
the (unavailable) transition probability governing the MJP. \cite{delmoral14} used a 
flexible Gaussian process to approximate the unavailable transition density of 
a diffusion process. Here, we take advantage of two well known continuous time 
approximations of the MJP by considering use of the transition density under a discretisation of the CLE or 
the tractable transition density under the LNA. The methods we propose are simple 
to implement and are not restricted to finite state spaces.

In section~2, a review of the basic structure of the problem is
presented, showing how the Markov process representation of a
reaction network is constructed. In section~3, we consider 
the problem of sampling conditioned MJPs and give three viable 
solutions to this problem. In section~4, it is shown how the recently
proposed particle MCMC algorithms \citep{andrieu09} may be applied to
this class of models. It is also shown how the bridge constructs introduced 
in section~3 can be used with a pMCMC scheme. The methodology is applied 
to a number of applications in section~5 before conclusions are drawn 
in section~6.

\section{Stochastic kinetic models}
\label{sec:stochkin}
We consider a reaction network involving $u$ species $\mathcal{X}_1,
\mathcal{X}_2,\ldots,\mathcal{X}_u$ and $v$ reactions $\mathcal{R}_1,\mathcal{R}_2,\ldots,\mathcal{R}_v$, 
with a typical reaction denoted by $\mathcal{R}_i$ and written using standard chemical reaction notation as
\begin{align*}
\mathcal{R}_i:\quad p_{i1}\mathcal{X}_1+p_{i2}\mathcal{X}_2+\cdots+p_{iu}\mathcal{X}_u &\longrightarrow
q_{i1}\mathcal{X}_1+q_{i2}\mathcal{X}_2+\cdots+q_{iu}\mathcal{X}_u 
\end{align*}
Let $X_{j,t}$ denote the number of molecules of species $\mathcal{X}_j$ at time
$t$, and let $X_t$ be the $u$-vector $X_t =
(X_{1,t},X_{2,t},\linebreak[1] \ldots,\linebreak[0] X_{u,t})'$. The dynamics of 
this model can be described by a vector of rates (or hazards) of the reactions together 
with a matrix which describes the effect of each reaction on the state. We therefore 
define a rate function $h_i(X_t,c_i)$, giving the overall hazard of a type $i$ reaction
occurring, and we let this depend explicitly on the reaction rate constant $c_i$, as well 
as the state of the system at time $t$. We model the system with a Markov jump process (MJP), 
so that for an infinitesimal time increment $dt$, the probability of a type $i$ reaction 
occurring in the time interval $(t,t+dt]$ is $h_i(X_t,c_i)dt$. When a type $i$ reaction 
does occur, the system state changes discretely, via the $i$th row of the so called 
net effect matrix $A$, a $v\times u$ matrix with $(i,j)$th element given by 
$q_{ij}-p_{ij}$. In what follows, for notational convenience, we work with the 
stoichiometry matrix defined as $S=A'$. Under the standard assumption of 
mass action kinetics, the hazard function for a particular 
reaction of type $i$ takes the form of the rate constant multiplied 
by a product of binomial coefficients expressing the number of ways in which 
the reaction can occur, that is,
\[
h_i(X_t,c_i) = c_i\prod_{j=1}^u \binom{X_{j,t}}{p_{ij}}.
\] 
Values for $c=(c_1,c_2,\ldots,c_v)'$ and the initial system state $X_0=x_0$
complete specification of the Markov process. Although this process is
rarely analytically tractable for interesting models, it is
straightforward to forward-simulate exact realisations of this Markov
process using a discrete event simulation method. This is due to the
fact that if the current time and state of the system are $t$ and
$X_t$ respectively, then the time to the next event will be
exponential with rate parameter
\[
h_0(X_t,c)=\sum_{i=1}^v h_i(X_t,c_i),
\]
and the event will be a reaction of type $\mathcal{R}_i$ with probability
$h_i(X_t,c_i)/h_0(X_t,c)$ independently of the waiting time. Forwards
simulation of process realisations in this way is typically referred
to as \emph{Gillespie's direct method} in the stochastic kinetics
literature, after \cite{Gillespie77}. See \cite{Wilkinson06} for
further background on stochastic kinetic modelling.
\index{Gillespie algorithm}

%In fact, the assumptions of mass-action kinetics, as well as the
%one-to-one correspondence between reactions and rate constants may
%both be relaxed. All of what follows is applicable to essentially
%arbitrary $v$-dimensional hazard functions $h(X_t,c)$.

The primary goal of this paper is that of inference for the stochastic 
rate constants $c$, given potentially noisy observations of the system state at a set of 
discrete times. \cite{GoliWilk11} demonstrated that it is possible to 
use a particle marginal Metropolis-Hastings (PMMH) scheme for 
such problems, using only the ability to forward simulate from the 
system of interest and evaluate the density associated with the 
observation error process. This ``likelihood free'' implementation 
uses the bootstrap particle filter of \cite{gordon93}. As noted by 
\cite{GoliWilk11}, the efficiency of this algorithm 
is likely to be very poor when observations are highly 
informative. Moreover, in the special case of error-free observation, 
the algorithm will be computationally intractable for models of 
realistic size and complexity. We therefore consider three constructions 
for generating realisations of conditioned jump processes, for use in 
a PMMH scheme. These constructs rely on the ability to construct both 
cheap and accurate approximations of the MJP. We therefore consider 
two candidate approximations in the next section.  

\subsection{SKM approximations}

\subsubsection{Chemical Langevin equation}

Over an infinitesimal time interval, $(t,t+dt]$, the
reaction hazards will remain constant almost surely. The
occurrence of reaction events can therefore be regarded as the
occurrence of events of a Poisson process with independent realisations
for each reaction type. Therefore, the mean and variance 
of the change in the MJP over the infinitesimal time interval 
can be calculated as 
\[
\operatorname{E}(dX_t)=S\,h(X_t,c)dt, \operatorname{Var}(dX_t)= S\operatorname{diag}\{h(X_t,c)\}S'dt.
\]
The It\^o stochastic differential equation (SDE) that has the same infinitesimal mean 
and variance as the true MJP is therefore
\begin{equation}
dX_t = S\,h(X_t,c)dt + \sqrt{S\operatorname{diag}\{h(X_t,c)\}S'}\,dW_t,
\label{cle}
\end{equation}
where (without loss of generality) $\sqrt{S\operatorname{diag}\{h(X_t,c)\}S'}$ 
is a $u\times u$ matrix and $W_t$ is a $u$-vector of standard Brownian 
motion. Equation \eqref{cle}
is the SDE most commonly referred to as the chemical Langevin
equation (CLE), and can be shown to approximate the SKM increasingly well in high
concentration scenarios \citep{Gillespie00}. The CLE can rarely be solved analytically, 
and it is common to work with a discretisation such as the Euler-Maruyama discretisation:
\[
\Delta X_t\equiv X_{t+\Delta t}-X_{t} = S\,h(X_t,c)\Delta t + \sqrt{S\operatorname{diag}\{h(X_t,c)\}S'\Delta t}\,Z
\]
where $Z$ is a standard multivariate Gaussian random variable.  

For a more formal discussion of the CLE 
and its derivation, we refer the reader to \cite{Gillespie92b} and
\cite{Gillespie00}.

\subsubsection{Linear noise approximation}

The linear noise approximation (LNA) further approximates the MJP by linearising 
the drift and noise terms of the CLE. The LNA generally possesses a greater degree of
numerical and analytic tractability than the CLE. For example, the LNA solution
involves (numerically) integrating a set of ODEs for which standard routines,
such as the \texttt{lsoda} package \citep{petzold83}, exist. The LNA can be derived in 
a number of more or less formal ways \citep{kurtz1970,elf2003,Komorowski09}. Our brief
derivation follows the approach of \cite{Wilkinson06} to which we refer the
reader for further details.

We begin by replacing the hazard function $h(X_t,c)$ in equation~(\ref{cle})
with the rescaled form $\Omega f(X_t /\Omega,c)$ where $\Omega$ is the volume of
the container in which the reactions are taking place. Note that the LNA
approximates the CLE increasingly well as $\Omega$ and $X_t$ become large, that
is, as the system approaches its thermodynamic limit. The CLE then becomes
\begin{equation}\label{lna1}
d X_t = \Omega S f(X_t /\Omega,c)dt + \sqrt{\Omega S\textrm{diag}\{f(X_t /\Omega,c)\}S'}\,dW_t.
\end{equation}
We now write the solution $X_t$ of the CLE as a
deterministic process plus a residual stochastic process \citep{kampen2001},
\begin{equation}\label{lna0}
X_t = \Omega z_{t}+\sqrt{\Omega}M_{t}.
\end{equation}
We then Taylor expand the rate function around $z_t$ to give
\begin{equation}\label{lna3}
f(z_t+M_t /\sqrt{\Omega},c) = f(z_t,c)+\frac{1}{\sqrt{\Omega}}F_t M_t + O(\Omega^{-1})
\end{equation}
where $F_t$ is the $v\times u$ Jacobian matrix with $(i,j)$th element $\partial f_{i}(z_t,c) / \partial z_{j,t}$ 
and we suppress the dependence of $F_t$ on $z_t$ and $c$ for simplicity. Substituting (\ref{lna0}) and (\ref{lna3}) 
into equation~(\ref{lna1}) and collecting terms of $O(1)$ and $O(1/\sqrt{\Omega})$ give the ODE satisfied by $z_t$, and SDE satisfied by $M_t$ 
respectively, as
\begin{align}
dz_{t}&=S\,f(z_{t},c)dt \label{lna4}\\
dM_{t}&=S\, F_t M_t dt + \sqrt{S\,\textrm{diag}\{f(z_t,c)\}S'}\,dW_t . \label{lna5}
\end{align}
Equations~(\ref{lna0}), (\ref{lna4}) and (\ref{lna5}) give the linear noise
approximation of the CLE and in turn, an approximation of the Markov jump process model.

For fixed or Gaussian initial conditions, that is $M_{0}\sim
\textrm{N}(m_{0},V_{0})$, the SDE in (\ref{lna5}) can be solved
explicitly to give
\[
M_{t}|c \sim \textrm{N}\left(G_{t}\, m_{0}\,,\,G_{t}\,\Psi_{t}\,G_{t}'\right)
\] 
where $G_t$ and $\Psi_t$ satisfy the coupled ODE system given by
\begin{align}
dG_{t}&= F_{t}G_{t}dt; \quad G_{0}=I_{u\times u},  \\
d\Psi_{t} &= G_{t}^{-1}S\,\textrm{diag}\{f(z_t,c)\}S' \left(G_{t}^{-1}\right)'; \quad \Psi_0 = V_0. 
\end{align}
Hence we obtain
\[
X_{t} \sim \textrm{N}\left(\Omega\,z_{t}+\sqrt{\Omega}\,G_{t}\, m_{0} \,,\, \Omega\,G_{t}\,\Psi_{t}\,G_{t}' \right).
\]  
In what follows we assume, without loss of generality, that $\Omega=1$.

\section{Sampling conditioned MJPs}
\label{sec:cond}
We suppose that interest lies in the Markov jump process 
over an interval $(0,t]$ denoted by $\mathbf{X}(t)=\{X_{s}\,|\, 0< s \leq t\}$.  
In fact, it is convenient to denote by $\mathbf{X}(t)$ the collection of reaction times and types 
over the interval $(0,t]$, which in turn gives the sample path of each species 
over this interval. Suppose further that the initial state $x_{0}$ is a known fixed value and that 
(a subset of components of) the 
process is observed at time $t$ subject to Gaussian error, giving a single observation 
$y_{t}$ on the random variable
\[
Y_{t}=P'X_{t}+\varepsilon_{t}\,,\qquad \varepsilon_{t}\sim \textrm{N}\left(0,\Sigma\right).
\]   
Here, $Y_{t}$ is a length-$p$ vector, $P$ is a constant matrix of dimension 
$u\times p$ and $\varepsilon_{t}$ is a length-$p$ Gaussian random vector. We denote the density 
linking $Y_{t}$ and $X_{t}$ as $p(y_{t}|x_{t})$. For simplicity, 
we assume in this section that both $\Sigma$ and the values of the rate constants $c$ are known, 
and drop them from the notation where possible. 

Our goal is to 
generate trajectories from $\mathbf{X}(t)|x_{0},y_{t}$ with associated probability function
\begin{align*}
\pi(\mathbf{x}(t)|x_{0},y_{t})&= \frac{p(y_{t}|x_{t})\pi(\mathbf{x}(t)|x_{0})}{p(y_{t}|x_{0})}\\
& \propto p(y_{t}|x_{t})\pi(\mathbf{x}(t)|x_{0})
\end{align*} 
where $\pi(\mathbf{x}(t)|x_{0})$ is the probability function associated 
with $\mathbf{x}(t)$. Although $\pi(\mathbf{x}(t)|x_{0},y_{t})$ will typically be intractable, 
simulation from $\pi(\mathbf{x}(t)|x_{0})$ is straightforward (via Gillespie's direct method), suggesting the construction of 
a numerical scheme such as Markov chain Monte Carlo or importance sampling. In keeping with 
the pMCMC methods described in section~4, we focus on the latter. 

\begin{algorithm}[t]
\caption{Myopic importance sampling}\label{mimp}
\begin{enumerate}
\item For $i=1,2,\ldots ,N$:
\begin{itemize}
\item[(a)] Draw $\mathbf{x}(t)^i\sim \pi(\mathbf{x}(t)|x_{0})$ using the Gillespie's direct method.
\item[(b)] Construct the unnormalised weight $\tilde{w}^{i}=p(y_{t}|x_{t}^i)$.
\end{itemize}
\item Normalise the weights: $w^{i}=\tilde{w}^{i} / \sum_{i=1}^{N}\tilde{w}^{i}$.
\item Resample (with replacement) from the discrete distribution on 
$\big\{\mathbf{x}(t)^1,\ldots,\mathbf{x}(t)^N\big\}$ using the normalised weights as probabilities.
\end{enumerate}
\end{algorithm}
The simplest importance 
sampling strategy (given in Algorithm~\ref{mimp}) proposes from $\pi(\mathbf{x}(t)|x_{0})$ 
and weights by $p(y_{t}|x_{t})$. If desired, an unweighted sample can easily be obtained 
by resampling (with replacement) from the discrete distribution over trajectory draws 
using the normalised weights as probabilities. Plainly, taking the average of the unnormalised 
weights gives an unbiased estimate of the normalising constant
\[
p(y_{t}|x_{0})=\textrm{E}_{\mathbf{X}(t)|x_{0}}\left(p(y_{t}|X_{t})\right).
\]  
This strategy is likely to work well provided that $y_{t}$ is not particularly 
informative. The proposal mechanism is independent of the observation $y_{t}$ 
and as $\Sigma$ is reduced, the variance of the importance weights increases. 
In an error free scenario, with $y_{t}\equiv x_{t}$, the unnormalised weights 
take the value 1 if $x_{t}^{i}=x_{t}$ and are 0 otherwise. Hence, in this 
extreme scenario, only trajectories that ``hit'' the observation have non-zero 
weight. 

In order to circumvent these problems, in section~\ref{sec:condhaz} 
we derive a novel proposal mechanism 
based on an approximation of the expected number of reaction events over 
the interval of interest, conditioned on the observation. In addition, in section~\ref{sec:bpf} 
we adapt a recently proposed bridge particle filter \citep{delmoral14} to our problem.        

%A simple but computational efficient 
%implementation is considered. In addition, we also adapt a recently proposed 
%bridge particle filter to our problem.       
\subsection{Conditioned hazard}
\label{sec:condhaz}

We suppose that we have simulated as far as time $s$ and derive 
an approximation of the expected number of reaction events over 
the interval $(s,t]$. Let $\Delta R_{s}$ denote the number of 
reaction events over the time $t-s=\Delta s$. We approximate $\Delta R_{s}$ 
by assuming a constant reaction hazard over $\Delta s$. A normal 
approximation to the corresponding Poisson distribution then gives
\[
\Delta R_{s}\sim \textrm{N}\left(h(x_s,c)\Delta s\,,\,H(x_s,c)\Delta s\right)
\]
where $H(x_s,c)=\textrm{diag}\{h(x_s,c)\}$. Under the Gaussian observation 
regime we have that
\[
Y_{t}|X_{s}=x_{s} \sim \textrm{N}\left(P'\left(x_{s}+S\,\Delta R_{s}\right)\,,\,P'S\,H(x_s,c)S'P\Delta s +\Sigma  \right).
\]
Hence, the joint distribution of $\Delta R_{s}$ 
and $Y_{t}$ can then be obtained approximately as
\[
\begin{pmatrix} \Delta R_{s} \\ Y_{t} \end{pmatrix}
\sim \textrm{N}\left\{\begin{pmatrix} h(x_s,c)\Delta s \\ P'\left(x_{s}+S\,h(x_s,c)\Delta s\right)\end{pmatrix}\,,\,
\begin{pmatrix} H(x_s,c)\Delta s & H(x_s,c)S'P\Delta s\\
P'S\,H(x_s,c)\Delta s & P'S\,H(x_s,c)S'P\Delta s +\Sigma\end{pmatrix}\right\}.
\]
Taking the expectation of $\Delta R_{s}|Y_{t}=y_{t}$ using standard multivariate normal theory, 
and dividing the resulting expression by $\Delta s$ gives 
an approximate conditioned hazard as
\begin{align}
&h^{*}(x_s,c|y_{t})=h(x_s,c) \nonumber \\
&\qquad+H(x_s,c)S'P\left(P'S\,H(x_s,c)S'P\Delta s +\Sigma\right)^{-1}\left(y_{t}-P'\left[x_{s}+S\,h(x_s,c)\Delta s\right]\right). \label{haz}
\end{align}
A proposed path $\mathbf{x}(t)^{*}$ can then be produced by sampling reaction events 
according to an inhomogeneous Poisson process with rate given by (\ref{haz}). An importance 
sampling scheme based on this proposal mechanism can then be obtained. Although the conditioned hazard in (\ref{haz}) 
depends on the current time $s$ in a nonlinear way, a simple implementation ignores 
this time dependence, giving exponential waiting times between proposed reaction events. 
Algorithm~\ref{condMJP} describes the mechanism for generating $\mathbf{x}(t)^{*}$.
\begin{algorithm}[t]
\caption{Approximate conditioned MJP generation}\label{condMJP}
\begin{enumerate}
\item Set $s=0$ and $x^{*}_{s}=x_{0}$. 
\item Calculate $h^{*}(x_{s}^{*},c|y_{t})$ and the combined hazard $h^{*}_{0}(x_{s}^{*},c|y_{t})=\sum_{i=1}^v h_{i}^{*}(x_{s}^{*},c_i|y_{t})$.
\item Simulate the time to the next event, $\tau\sim \textrm{Exp}\{h^{*}_{0}(x_{s}^{*},c|y_{t})\}$.
\item Simulate the reaction index, $j$, as a discrete random quantity with
  probabilities proportional to $h_{i}^{*}(x_{s}^{*},c_{i}|y_{t})$, $i=1,\ldots ,v$.
\item Put $x_{s+\tau}^{*}=x_{s}+S^{j}$ where $S^{j}$ denotes the $j$th column of $S$. Put $s:=s+\tau$.
\item Output $x_{s}^{*}$ and $s$. If $s<t$, return to step 2.
\end{enumerate}
\end{algorithm}
To calculate the importance weights, we first note that $\pi(\mathbf{x}(t)|x_{0})$ can be 
written explicitly by considering the generation of all reaction times and types over 
$(0,t]$. To this end, we let $r_{j}$ denote the number of reaction events of type $\mathcal{R}_{j}$, 
$j=1,\ldots,v$, and define $n_{r}=\sum_{j=1}^{v}r_{j}$ as the total number of reaction events 
over the interval. Reaction times (assumed to be in increasing order) and types are denoted by 
$(t_{i},\nu_{i})$, $i=1,\ldots ,n_{r}$, $\nu_{i}\in \{1,\ldots ,v\}$ and we take $t_{0}=0$ and 
$t_{n_{r}+1}=t$. \cite{Wilkinson06} gives $\pi(\mathbf{x}(t)|x_{0})$, also known as the complete 
data likelihood over $(0,t]$, as
\begin{align*}
\pi(\mathbf{x}(t)|x_{0})&=\left\{\prod_{i=1}^{n_{r}}h_{\nu_{i}}\left(x_{t_{i-1}},c_{\nu_{i}}\right)\right\}
\exp\left\{-\sum_{i=1}^{n_{r}}h_{0}\left(x_{t_{i}},c\right)\left[t_{i+1}-t_{i}\right]\right\}\\
&= \left\{\prod_{i=1}^{n_{r}}h_{\nu_{i}}\left(x_{t_{i-1}},c_{\nu_{i}}\right)\right\}
\exp\left\{-\int_{0}^{t}h_{0}\left(x_{t},c\right)\,dt\right\}.
\end{align*} 
We let $q(\mathbf{x}(t)|x_{0},y_{t})$ denote the complete data likelihood under the 
approximate jump process with hazard $h^{*}(x_{s}^{*},c|y_{t})$, so that 
the importance weight for a path $\mathbf{x}(t)$ is given by
\begin{align}
&p(y_{t}|x_{t})\frac{\pi(\mathbf{x}(t)|x_{0})}{q(\mathbf{x}(t)|x_{0},y_{t})}\nonumber \\
&\qquad =p(y_{t}|x_{t})\left\{\prod_{i=1}^{n_{r}}\frac{h_{\nu_{i}}\left(x_{t_{i-1}},c_{\nu_{i}}\right)}{h^{*}_{\nu_{i}}\left(x_{t_{i-1}},c_{\nu_{i}}|y_{t}\right)}\right\}
\exp\left\{-\sum_{i=1}^{n_{r}}\left[h_{0}\left(x_{t_{i}},c\right)-h^{*}_{0}\left(x_{t_{i}},c|y_{t}\right)\right]\left[t_{i+1}-t_{i}\right]\right\}. \label{weight}
\end{align}
When the inhomogeneous Poisson  process approximation is sampled exactly, the importance weight in (\ref{weight}) 
becomes
\begin{align*}
& p(y_{t}|x_{t})\left\{\prod_{i=1}^{n_{r}}\frac{h_{\nu_{i}}\left(x_{t_{i-1}},c_{\nu_{i}}\right)}{h^{*}_{\nu_{i}}\left(x_{t_{i-1}},c_{\nu_{i}}|y_{t}\right)}\right\}
\exp\left\{-\int_{0}^{t}\left[h_{0}\left(x_{t},c\right)-h^{*}_{0}\left(x_{t},c|y_{t}\right)\right]\,dt\right\}\\
&\qquad =p(y_{t}|x_{t})\frac{d\mathbb{P}}{d\mathbb{Q}}\left(\mathbf{x}(t)\right)
\end{align*}
where the last term is seen to be the Radon-Nikodym derivative of the true Markov jump process 
($\mathbb{P}$) with respect to the inhomogeneous Poisson process approximation ($\mathbb{Q}$) 
and measures the closeness of the approximating process to the true process.

Algorithm~\ref{impCondMJP} gives importance sampling algorithm that uses an approximate implementation 
of the inhomogeneous Poisson process approximation. Note that in the 
special case of no error, the importance weight in step 1(b) has $p(y_{t}|x_{t}^{i})$ replaced with an indicator function 
assigning the value 1 if $x_{t}^{i}=x_{t}$ and 0 otherwise. Upon completion of the algorithm, an equally 
weighted sample approximately distributed according to $\pi(\mathbf{x}(t)|x_{0},y_{t})$ is obtained. The 
average unnormalised weight can be used to (unbiasedly) estimate the normalising constant $p(y_{t}|x_{0})$.
\begin{algorithm}[t]
\caption{Importance sampling with conditioned hazard}\label{impCondMJP}
\begin{enumerate}
\item For $i=1,2,\ldots ,N$:
\begin{itemize}
\item[(a)] Draw $\mathbf{x}(t)^i\sim q(\mathbf{x}(t)|x_{0},y_{t})$ using Algorithm~\ref{condMJP}.
\item[(b)] Construct the unnormalised weight 
\[
\tilde{w}^{i}=p(y_{t}|x_{t}^{i})\frac{\pi(\mathbf{x}(t)^{i}|x_{0})}{q(\mathbf{x}(t)^{i}|x_{0},y_{t})}
\]
whose form is given by (\ref{weight}).
\end{itemize}
\item Normalise the weights: $w^{i}=\tilde{w}^{i} / \sum_{i=1}^{N}\tilde{w}^{i}$.
\item Resample (with replacement) from the discrete distribution on 
$\big\{\mathbf{x}(t)^1,\ldots,\mathbf{x}(t)^N\big\}$ using the normalised weights as probabilities.
\end{enumerate}
\end{algorithm}

\subsection{Bridge particle filter}
\label{sec:bpf}

\cite{delmoral14} considered the problem of sampling continuous time, continuous 
valued Markov processes and proposed a bridge particle filter to weight forward 
trajectories based on an approximation to the unknown transition probabilities 
at each reweighting step. Here, we adapt their method to our problem. We note that 
when using the bridge particle filter to sample MJP trajectories, it is possible 
to obtain a likelihood free scheme.

Without loss of generality, we adopt an equispaced partition of $[0,t]$ as 
\[
0=t_{0}<t_{1}<\cdots < t_{n}=t.
\]  
This partition is used to determine the times at which resampling may take place. 
Introduce the weight functions
\[
W_{k}(x_{t_{k-1}:t_{k}})=\frac{q(y_{t}|x_{t_{k}})}{q(y_{t}|x_{t_{k-1}})}
\]  
where $q(y_{t}|x_{t_{k}})$, $k=0,\ldots ,n$ are positive functions. Note that
\[
\frac{q(y_{t}|x_{0})}{q(y_{t}|x_{t})}\prod_{k=1}^{n}W_{k}(x_{t_{k-1}:t_{k}})=1
\]
and write $\pi(\mathbf{x}(t)|x_{0},y_{t})$ as
\begin{align}
\pi(\mathbf{x}(t)|x_{0},y_{t})&\propto p(y_{t}|x_{t})\pi(\mathbf{x}(t)|x_{0})\frac{q(y_{t}|x_{0})}{q(y_{t}|x_{t})}\prod_{k=1}^{n}W_{k}(x_{t_{k-1}:t_{k}})\nonumber \\
&\propto p(y_{t}|x_{t})\frac{q(y_{t}|x_{0})}{q(y_{t}|x_{t})}\prod_{k=1}^{n}W_{k}(x_{t_{k-1}:t_{k}})\pi(\mathbf{x}(t_{k-1}:t_{k})|x_{t_{k-1}})\nonumber \\
&\propto \prod_{k=1}^{n}W_{k}(x_{t_{k-1}:t_{k}})\pi(\mathbf{x}(t_{k-1}:t_{k})|x_{t_{k-1}}) \label{target}
\end{align} 
where $\pi(\mathbf{x}(t_{k-1}:t_{k})|x_{t_{k-1}})$ denotes the probability function associated 
with $\mathbf{X}(t_{k-1}:t_{k})=\{X_{s}\,|\, t_{k-1}< s \leq t_{k}\}$ and the last line (\ref{target}) 
follows by taking $q(y_{t}|x_{t})$ to be $p(y_{t}|x_{t})$ and absorbing $q(y_{t}|x_{0})$ 
into the proportionality constant. The form of (\ref{target}) suggests a sequential 
Monte Carlo (SMC) scheme (also known as a particle filter) where at time $t_{k-1}$ 
each particle (trajectory) $\mathbf{x}(t_{k-1})^{i}$ is extended by simulating from 
$\pi(\mathbf{x}(t_{k-1}:t_{k})|x_{t_{k-1}}^{i})$ and incrementally weighted by 
$W_{k}(x_{t_{k-1}:t_{k}})$. Intuitively, by ``looking ahead'' to the observation, 
trajectories that are not consistent with $y_{t}$ are given small weight and 
should be pruned out with a resampling step. \cite{delmoral14} suggest an adaptive 
resampling procedure so that resampling is only performed if the effective 
sample size (ESS) falls below some fraction of the number of particles, say $\beta$. 
The ESS is defined \citep{liu1995} as a function of the weights $w^{1:N}$ by
\begin{equation}\label{ess}
ESS\left(w^{1:N}\right)=\frac{\left(\sum_{i=1}^{N}w^{i}\right)^{2}}{\sum_{i=1}^{N}\left(w^{i}\right)^{2}}\,.
\end{equation}

It remains that we can choose sensible functions $q(y_{t}|x_{t_{k}})$ to be used 
to construct the weights. We propose to use the density associated with 
$Y_{t}|X_{t_{k}}=x_{t_{k}}$ under the CLE or LNA:
\begin{align*}
q_{CLE}(y_{t}|x_{t_{k}})&=\textrm{N}\left(y_{t};\,P'\left[x_{t_{k}}+S\,h(x_{t_{k}},c)(t-t_{k})\right]\,,\,P'S\,H(x_{t_{k}},c)S'P(t-t_{k})+\Sigma\right),\\
q_{LNA}(y_{t}|x_{t_{k}})&=\textrm{N}\left(y_{t};\,P'\left[z_{t}+G_{t-t_{k}}\, (x_{t_{k}}-z_{t_{k}})\right] \,,\, P'G_{t-t_{k}}\,\Psi_{t-t_{k}}\,G_{t-t_{k}}P+\Sigma\right).
\end{align*}
Note that due to the intractability of the CLE, we propose to use a single step of the Euler-Mauyama approximation. 
Comments on the relative merits of each scheme are given in Section~\ref{sec:eff}. 

Algorithm~\ref{bridgePF} gives the sequence of steps necessary to implement the bridge particle filter. 
The average unnormalised weight obtained at time $t$ can be used to estimate 
the normalising constant $p(y_{t}|x_{0})$:
\[
\widehat{p}(y_{t}|x_{0})\propto \frac{1}{N}\sum_{i=1}^{N}\tilde{w}_{n}^{i}\,.
\] 
\begin{algorithm}[t]
\caption{Bridge particle filter}\label{bridgePF}
\begin{enumerate}
\item Initialise. For $i=1,2,\ldots ,N$:
\begin{itemize}
\item[(a)] Set $x_{0}^{i}=x_{0}$ and put $w_{0}^{i}=1/N$.
\end{itemize}
\item Perform sequential importance sampling. For $k=1,2,\ldots ,n$ and 
$i=1,2,\ldots ,N$: 
\begin{itemize}
\item[(a)] If $ESS(w_{k-1}^{1:N})<\beta N$ draw indices $a_{k}^{i}$ from the discrete 
distribution on $\{1,\ldots,N\}$ with probabilities given by $w_{k-1}^{1:N}$ and put 
$w_{k}^{i}=1/N$. Otherwise, put $a_{k}^{i}=i$.
\item[(b)] Draw $\mathbf{x}(t_{k-1}:t_{k})^{i}\sim \pi(\cdot|x_{t_{k-1}}^{a_{k}^{i}})$ using the Gillespie algorithm initialised 
at $x_{t_{k-1}}^{a_{k}^{i}}$.
\item[(c)] Construct the unnormalised weight 
\[
\tilde{w}_{k}^{i}=\tilde{w}_{k-1}^{i}\frac{q(y_{t}|x_{t_{k}}^{i})}{q(y_{t}|x_{t_{k-1}}^{a_{k}^{i}})}
\]
\item[(d)] Normalise the weights: $w_{k}^{i}=\tilde{w}_{k}^{i} / \sum_{i=1}^{N}\tilde{w}_{k}^{i}$.
\end{itemize}
\item Let $b_{n}^{i}=i$ and define $b_{k}^{i}=a_{k+1}^{b_{k+1}^{i}}$ recursively. Resample (with replacement) 
from the discrete distribution on $\big\{(\mathbf{x}(0:t_{1})^{b_{1}^{i}},\ldots,\mathbf{x}(t_{n-1}:t)^{i}),i=1,\ldots,N\big\}$ 
using the normalised weights as probabilities.
\end{enumerate}
\end{algorithm}

We now consider some special cases of Algorithm~\ref{bridgePF}. For unknown $x_{0}$ with 
prior probability mass function $\pi(x_{0})$, the target becomes
\[
\pi(\mathbf{x}(t),x_{0}|y_{t})\propto \pi(x_{0})q(y_{t}|x_{0})\prod_{k=1}^{n}W_{k}(x_{t_{k-1}:t_{k}})\pi(\mathbf{x}(t_{k-1}:t_{k})|x_{t_{k-1}})
\] 
which suggests that step 1(a) should be replaced by sampling particles $x_{0}^{i}\sim \pi(x_{0})$. 
The contribution $q(y_{t}|x_{0})$ could either be absorbed into the final weight (taking care to respect the 
ancestral lineage of the trajectory), or an initial look ahead step could be performed by resampling 
amongst the $x_{0}^{i}$ with weights proportional to $q(y_{t}|x_{0}^{i})$. If the latter strategy is adopted 
and no additional resampling steps are performed, the algorithm reduces to the auxiliary particle filter 
of \cite{pitt1999}, where particles are pre-weighted by $q(y_{t}|x_{0})$ and propagated through myopic forward 
simulation. If no resampling steps are performed at any time, the algorithm reduces to the myopic importance 
sampling strategy described in Section~\ref{mimp}.

In the error free scenario, the target can be written as 
\[
\pi(\mathbf{x}(t)|x_{t},x_{0})\propto\frac{q(x_{t}|x_{0})}{q(x_{t}|x_{t_{n-1}})} \pi(\mathbf{x}(t_{n-1}:t)|x_{t_{n-1}}) \prod_{k=1}^{n-1}W_{k}(x_{t_{k-1}:t_{k}})\pi(\mathbf{x}(t_{k-1}:t_{k})|x_{t_{k-1}})
\]
where the incremental weight functions are redefined as
\[
W_{k}(x_{t_{k-1}:t_{k}})=\frac{q(x_{t}|x_{t_{k}})}{q(x_{t}|x_{t_{k-1}})}\,.
\]
The form of the target suggests that at time $t_{n-1}$, particle trajectories should 
be propagated via $\pi(\mathbf{x}(t_{n-1}:t)|x_{t_{n-1}})$ and weighted by 
$q(x_{t}|x_{0})/q(x_{t}|x_{t_{n-1}})$, provided that the trajectory ``hits'' 
the observation $x_t$, otherwise a weight of 0 should be assigned. Hence, unlike 
in the continuous state space scenario considered by \cite{delmoral14}, the 
algorithm is likelihood free, in the sense that $\pi(\mathbf{x}(t_{n-1}:t)|x_{t_{n-1}})$ 
need not be evaluated. 

\subsection{Comments on efficiency}
\label{sec:eff}

Application of Algorithm~\ref{impCondMJP} requires calculation of the conditioned 
hazard function in (\ref{haz}) after every reaction event. The cost of this 
calculation will therefore be dictated by the number of observed components 
$p$, given that a $p\times p$ matrix must be inverted. Despite this, for many 
systems of interest, it is unlikely that all components will be 
observed and we anticipate that in practice $p< <u$, where $u$ is the number of species in the 
system. The construction of the conditioned hazard is based 
on an assumption that the hazard function is constant over diminishing 
time intervals $(s,t]$ and that the number of reactions over this interval 
is approximately Gaussian. The performance of the construct is therefore 
likely to diminish if applied over time horizons during which the 
reaction hazards vary substantially. We also note that the elements of the conditioned hazard 
are not guaranteed to be positive and we therefore truncate each hazard component at zero.  

We implement the bridge particle filter in Algorithm~\ref{bridgePF} 
with the weight functions obtained either through the CLE or the LNA. To 
maximise statistical efficiency, we require that $q(y_{t}|x_{t_{k}})\approx 
p(y_{t}|x_{t_{k}})$. Given the analytic intractability of the CLE, we 
obtain $q$ via a single step of an Euler-Maruyama scheme. Whilst this 
is likely to be computationally efficient, given the simplicity 
of applying a single step of the Euler-Maruyama scheme, we anticipate 
that applying the scheme over large time intervals (where non-linear dynamics 
are observed) is likely to be unsatisfactory. The tractability of the LNA has been recently exploited 
\citep{Komorowski09,fearnhead12,Goli14} and shown to give a reasonable approximation to the MJP 
for a number of reaction networks. However, use of the LNA requires the solution of a system of $u(u+1)/2 +2u$ 
coupled ODEs. For most stochastic kinetic models of interest, the solution 
to the LNA ODEs will not be analytically tractable. Whilst good numerical ODE solvers 
are readily available, the bridge particle filter is likely to require a 
full numerical solution over the time interval of interest for each particle (except 
in the special error free case where only a single solution is required). 
Both the CLE and LNA replace the intractable transition probability with a 
Gaussian approximation. Moreover, the approximations may be light tailed relative to the 
target, and erstwhile valid trajectories may be pruned out by the resampling 
procedure. Tempering the approximations by raising $q(y_t|x_{t_{k}})$ to a power 
$\gamma$ ($0<\gamma<1$) may alleviate this problem at the expense of choosing an appropriate value 
for the additional tuning parameter $\gamma$. We assess the empirical performance 
of each scheme in Section~\ref{sec:app}.   

\section{Bayesian inference}
\label{sec:bayes}

Consider a time interval $[0,T]$ over which a Markov jump process $\mathbf{X}=\{X_{t}\,|\, 0\leq t \leq T\}$ 
is not observed directly, but 
observations (on a regular grid) $\mathbf{y}=\{y_{t}\,|\, t=0,1,\ldots ,T\}$ are available 
and assumed conditionally independent (given $\mathbf{X}$) with conditional probability 
distribution obtained via the observation equation,
\begin{equation}\label{obs_eq}
Y_{t}=P'X_{t}+\varepsilon_{t},\qquad \varepsilon_{t}\sim \textrm{N}\left(0,\Sigma\right),\qquad t=0,1,\ldots ,T.
\end{equation} 
As in Section~\ref{sec:cond}, we take $Y_{t}$ to be a length-$p$ vector, $P$ is a constant matrix of dimension 
$u\times p$ and $\varepsilon_{t}$ is a length-$p$ 
Gaussian random vector. We assume that primary interest lies in the rate constants $c$ where, in the case of unknown 
measurement error variance, the parameter vector $c$ is augmented to include the parameters of $\Sigma$. 
Bayesian inference may then proceed through the marginal posterior density
%\begin{eqnarray}
%p(c,\mathbf{x}|\mathbf{y})&\propto& p(c)\,p(\mathbf{x}|c)\,p(\mathbf{y}|\mathbf{x},c)\nonumber \\
%			      &\propto& p(c)\,p(\mathbf{x}|c)\,\prod_{t=1}^{T}p(y_{t}|x_{t},c)\label{jp}
%\end{eqnarray}
%\begin{eqnarray}
%\pi(c,\mathbf{x}|\mathbf{y})&\propto& \pi(c)\,\pi(\mathbf{x}|c)\,\prod_{t=0}^{T}p(y_{t}|x_{t},c)\label{jp}
%\end{eqnarray}
\begin{equation}
p(c|\mathbf{y})\propto p(c)p(\mathbf{y}|c)\label{jp}
\end{equation}
where $p(c)$ is the prior density ascribed to $c$ and 
$p(\mathbf{y}|c)$ is the marginal likelihood. Since the posterior 
in (\ref{jp}) will be intractable in practice, samples are usually 
generated from (\ref{jp}) via MCMC. A further complication is the 
intractability of the marginal likelihood term, and we therefore 
adopt the particle marginal Metropolis-Hastings scheme of 
\cite{andrieu10} which has been successfully applied to stochastic 
kinetic models in \cite{GoliWilk11} and \cite{Goli14}.

\subsection{Particle marginal Metropolis-Hastings}
\label{sec:pmmh}

Since interest lies in the marginal posterior in (\ref{jp}), 
we consider the special case of the particle marginal Metropolis-Hastings 
(PMMH) algorithm \citep{andrieu10} which can be seen as a pseudo-marginal 
MH scheme \citep{beaumont03,andrieu09b}. Under some fairly mild conditions (for which 
we refer the reader to \cite{delmoral04} and \cite{andrieu10}), a sequential Monte Carlo scheme 
targeting the probability associated with the conditioned MJP, $\pi(\mathbf{x}|\mathbf{y},c)$, 
can be implemented to give an unbiased estimate of the marginal likelihood. 
We write this estimate as $\widehat{p}(\mathbf{y}|c,u)$ where $u$ denotes all 
random variables generated by the SMC scheme according to some density $q(u|c)$. 
We now consider a target of the form
\[
\widehat{p}(c,u|\mathbf{y})\propto \widehat{p}(\mathbf{y}|c,u)q(u|c)p(c)
\]
for which marginalisation over $u$ gives
\begin{align*}
\int\widehat{p}(c,u|\mathbf{y})\,du&\propto p(c)\textrm{E}_{u|c}\left\{\widehat{p}(\mathbf{y}|c,u)\right\}\\
&\propto p(c)p(\mathbf{y}|c).
\end{align*}
Hence, a MH scheme targeting $\widehat{p}(c,u|\mathbf{y})$ 
with proposal kernel $q(c^{*}|c)q(u^{*}|c^{*})$ accepts a move from $(c,u)$ 
to $(c^{*},u^{*})$ with probability
\[
\frac{\widehat{p}\big(\mathbf{y}|c^{*},u^{*}\big)p\big(c^{*}\big)}{\widehat{p}\big(\mathbf{y}|c,u\big)p\big(c\big)} \times \frac{q\big(c|c^{*}\big)}
{q\big(c^{*}|c\big)}.
\]
We see that the values of $u$ need never be stored and it should now be clear 
that the scheme targets the correct marginal distribution $p(c|\mathbf{y})$.

\subsection{Implementation}
\label{sec:implement}

Algorithms~\ref{impCondMJP} and \ref{bridgePF} can readily be applied to 
give an SMC scheme targeting $\pi(\mathbf{x}|\mathbf{y},c)$. In each case, 
an initialisation step should be performed where a weighted sample 
$\{(x_{0}^{i},w_{0}^{i}),i=1,\ldots ,N\}$ is obtained by drawing values 
$x_{0}^{i}$ from some prior with mass function $\pi(x_0)$ and assigning 
weights proportional to $p(y_{0}|x_{0}^{i},c)$. If desired, resampling 
could be performed so that the algorithm is initialised with an equally 
weighted sample drawn from $\pi(x_{0}|y_{0},c)$. Algorithms~\ref{impCondMJP} 
and \ref{bridgePF} can then be applied sequentially, for times 
$t=1,2,\ldots,T$, simply by replacing $x_{0}$ with $x_{t-1}^{i}$. After 
assimilating all information, an unbiased estimate of the marginal likelihood 
$p(\mathbf{y}|c)$ is obtained as
\begin{equation}\label{margll}
\widehat{p}(\mathbf{y}|c)=\widehat{p}(y_{0}|c)\prod_{t=1}^{T}\widehat{p}(y_{t}|\mathbf{y}(t-1))
\end{equation}
where $\mathbf{y}(t-1)=\{y_{t},t=0,1,\ldots,t-1\}$ and we have dropped 
$u$ from the notation for simplicity. The product in (\ref{margll}) can be obtained 
from the output of Algorithms~\ref{impCondMJP} and \ref{bridgePF}. For example, 
when using the conditioned hazard approach, (\ref{margll}) is simply the product 
of the average unnormalised weight obtained in step 1(b). Use of 
Algorithms~\ref{impCondMJP} and \ref{bridgePF} in this way give SMC schemes 
that fall into a class of auxiliary particle filters \citep{pitt1999}. We refer the 
reader to \cite{pitt12} for a theoretical treatment of the use of an auxiliary particle 
filter inside an MH scheme.   

The mixing of the PMMH scheme is likely to depend on the number of particles used in the SMC scheme. Whilst 
the method can be implemented using just $N=1$ particle, the corresponding 
estimator of marginal likelihood will be highly variable, and the impact 
of this on the PMMH algorithm will be a poorly mixing chain. As noted by 
\cite{andrieu09b}, the mixing efficiency of the PMMH 
scheme decreases as the variance of the estimated marginal likelihood 
increases. This problem can be alleviated at the expense of greater 
computational cost by increasing $N$. This therefore suggests an optimal 
value of $N$ and finding this choice is the subject of \cite{sherlock2013} 
and \cite{doucet13}. The former show that for a ``standard asymptotic regime'' 
$N$ should be chosen so that the variance in the noise in the estimated 
log-posterior is around 3, but find that for low dimensional problems a 
smaller value (around 2) is optimal.  We therefore recommend 
performing an initial pilot run of PMMH to obtain an estimate of the 
posterior mean (or median) parameter value, and a (small) number of additional 
sampled values. The value of $N$ should then be chosen so that the variance 
of the noise in the estimated log-posterior is (ideally) in the range $[2,4]$.

Since all parameter values must be strictly positive we adopt a proposal kernel 
corresponding to a random walk on $\log(c)$, with Gaussian innovations. We 
take the innovation variance to be $\lambda \widehat{\textrm{Var}}(\log(c))$ 
and follow the practical advice of \cite{sherlock2013} by tuning $\lambda$ 
to give an acceptance rate of around 15\%. 

\section{Applications}
\label{sec:app}
In order to examine the empirical performance of the methods proposed 
in section~\ref{sec:cond}, we consider three examples. These are a 
simple (and tractable) birth-death model, the stochastic Lotka-Volterra 
model examined by \citet{BWK08} and a systems biology model of bacterial motility regulation 
\citep{Wilkinson11}.

\subsection{Birth-Death}
\label{sec:bdmodel}

The birth-death reaction network takes the form
\[
\mathcal{R}_{1}:\, \mathcal{X}_{1} \longrightarrow 2\mathcal{X}_{1},\quad
\mathcal{R}_{2}:\,\mathcal{X}_{1} \longrightarrow \emptyset
\]
with birth and death reactions shown respectively. The stoichiometry matrix is given by
\[
S = \begin{pmatrix} 
1 & -1 
\end{pmatrix}
\]
and the associated hazard function is 
\[
h(x_t,c) = (c_{1}x_{t}, c_{2}x_{t})'
\]
where $x_t$ denotes the state of the system at time $t$. The CLE is given by
\[
dX_{t}=\left(c_{1}-c_{2}\right)X_{t}\,dt + \sqrt{\left(c_{1}+c_{2}\right)X_{t}}\,dW_{t}
\]
which can be seen as a degenerate case of a Feller square-root diffusion \citep{feller52}. 
For reaction networks of reasonable size and complexity, the CLE will be intractable. 
To explore the effect of working with a numerical approximation of the CLE 
inside the bridge particle filter, we adopt the Euler-Maruyama approximation which gives 
(for a fixed initial condition $x_0$) an approximation to the transition density as
\[
X_{t}|X_{0}=x_{0}\sim \textrm{N}\left(x_{0}+\left(c_{1}-c_{2}\right)x_{0}\,t\,,\,\left(c_{1}+c_{2}\right)x_{0}\,t\right).
\] 
The ODE system governing the LNA with initial conditions $z_{0}=x_{0}$, $m_{0}=0$ and $V_{0}=0$ 
can be solved analytically to give
\[
X_{t}|X_{0}=x_{0}\sim \textrm{N}\left(x_{0} e^{(c_{1}-c_{2})t}\,,\,x_{0}\frac{(c_{1}+c_{2})}{(c_{1}-c_{2})} e^{(c_{1}-c_{2})t} \left[e^{(c_{1}-c_{2})t} - 1\right]\right).
\]
We consider a an example in which $c=(0.5,1)$ and $x_{0}=100$ are fixed. To provide a challenging scenario 
we took $x_{t}$ to be the upper 99\% quantile of $X_{t}|X_{0}=100$. To assess the performance of each algorithm 
as an observation is made with increasing time sparsity, we took $t\in\{0.1,0.5,1\}$. Algorithms~\ref{mimp} (denoted MIS), 
\ref{impCondMJP} (denoted CH) and \ref{bridgePF} (denoted BPF-CLE or BPF-LNA) were run with $N\in\{10,50,100,500\}$ to give a set of $m=5000$ 
estimates of the transition probability $\pi(x_{t}|x_{0})$ and we denote this set by $\widehat{\pi}^{1:m}_{N}(x_{t}|x_{0})$. 
The bridge particle filter also requires specification of the intermediate time points at which 
resampling could take place. For simplicity, we took an equispaced partition of $[0,t]$ with 
a time step of 0.02 for $t=0.1$, and $0.05$ for $t\in\{0.5,1\}$. We found that these gave 
a good balance between statistical efficiency and CPU time.    

\begin{table}[t]
	\centering
	\begin{tabular}{|l|c|c|c|c|}
	\hline	
Method	& $N$ & $t=0.1$ & $t=0.5$ & $t=1$ \\
	\hline
MIS & 10  & 300, 293, 6.2$\times 10^{-4}$ &171, 168, 3.5$\times 10^{-4}$ &151, 149, 3.0$\times 10^{-4}$  \\
    & 50  &1340, 1190, 1.2$\times 10^{-4}$ &827, 773, 7.0$\times 10^{-5}$ &682, 639, 5.8$\times 10^{-5}$  \\
    & 100 &2331, 1921, 6.4$\times 10^{-5}$ &1488, 1308, 3.5$\times 10^{-5}$ &1364, 1203, 3.2$\times 10^{-5}$  \\
    & 500 &4776, 3771, 1.2$\times 10^{-5}$ &4196, 3230, 6.8$\times 10^{-6}$ &3901, 3004, 6.1$\times 10^{-6}$  \\
\hline
CH  & 10  &4974, 3264, 1.6$\times 10^{-5}$ &4985, 2998, 7.8$\times 10^{-6}$ &4990, 3581, 2.4$\times 10^{-6}$  \\
    & 50  &5000, 4395, 4.6$\times 10^{-6}$ &5000, 4546, 1.2$\times 10^{-6}$ &5000, 4508, 9.7$\times 10^{-7}$  \\
    & 100 &5000, 4689, 2.4$\times 10^{-6}$ &5000, 4668, 8.5$\times 10^{-7}$ &5000, 4798, 3.8$\times 10^{-7}$  \\
    & 500 &5000, 4921, 7.7$\times 10^{-7}$ &5000, 4943, 1.6$\times 10^{-7}$ &5000, 4939, 1.2$\times 10^{-7}$  \\
\hline
BPF-CLE & 10  &2581, 349, 5.7$\times 10^{-4}$ &2412, 556, 7.7$\times 10^{-5}$ &2745, 532, $1.7\times 10^{-5}$  \\
        & 50  &4982, 2137, 6.3$\times 10^{-5}$ &4920, 3391, 4.9$\times 10^{-6}$ &3236, 4925, 4.0$\times 10^{-6}$ \\
        & 100 &5000, 3519, 1.9$\times 10^{-5}$ &4998, 3979, 2.8$\times 10^{-6}$ &4999, 4106, 4.1$\times 10^{-6}$   \\
        & 500 &5000, 3841, 1.5$\times 10^{-5}$ &5000, 4756, 6.7$\times 10^{-7}$ &5000, 4780, 3.2$\times 10^{-6}$  \\
\hline
BPF-LNA & 10  &2634, 403, 4.3$\times 10^{-4}$ &2514, 636, 6.9$\times 10^{-5}$ &2843, 1102, 2.4$\times 10^{-5}$  \\
        & 50  &4963, 2748, 3.2$\times 10^{-5}$ &4926, 3198, 6.0$\times 10^{-6}$ &4949, 3625, 2.8$\times 10^{-6}$  \\
        & 100 &5000, 3612, 1.5$\times 10^{-5}$ &4998, 4055, 2.5$\times10^{-6}$ &5000, 4016, 1.9$\times 10^{-6}$  \\
        & 500 &5000, 3643, 1.4$\times 10^{-5}$ &5000, 4655, 8.8$\times 10^{-7}$ &5000, 4771, 5.4$\times 10^{-7}$  \\
	\hline
	\end{tabular}
	\caption{$\sum_{i=1}^{m}I(\widehat{\pi}_{N}(x_{t}|x_{0})>0)$, $\textrm{ESS}(\widehat{\pi}^{1:m}_{N}(x_{t}|x_{0}))$ and 
$\textrm{MSE}(\widehat{\pi}^{1:m}_{N}(x_{t}|x_{0}))$, based on 5000 runs of MIS, CH, BPF-CLE and BPF-LNA. For MIS, 
the expected number of non-zero estimates (as obtained analytically) is reported. 
In all cases, $x_{0}=100$ and $x_{t}$ is the upper 99\% quantile of $X_{t}|X_{0}=100$.}\label{tab:tabBD}
\end{table}

To compare the algorithms, we report the number of non-zero normalising constant estimates 
$\sum_{i=1}^{m}I(\widehat{\pi}_{N}(x_{t}|x_{0})>0)$, the effective sample size 
$\textrm{ESS}(\widehat{\pi}^{1:m}_{N}(x_{t}|x_{0}))$ whose form is defined in (\ref{ess}) and 
mean-squared error $\textrm{MSE}(\widehat{\pi}^{1:m}_{N}(x_{t}|x_{0}))$ given by
\[
\textrm{MSE}(\widehat{\pi}^{1:m}_{N}(x_{t}|x_{0}))=\frac{1}{m}\sum_{i=1}^{m}
\left[\widehat{\pi}^{i}_{N}(x_{t}|x_{0})-\pi(x_{t}|x_{0})\right]^{2}
\]
where $\pi(x_{t}|x_{0})$ can be obtained analytically \citep{bailey1964}.

The results are summarised in Table~\ref{tab:tabBD}. Use of the 
conditioned hazard and bridge particle filters (CH, BPF-CLE and BPF-LNA) comprehensively 
outperform the myopic importance sampler (MIS). For example, for the $t=1$ case, an 
order of magnitude improvement is observed when comparing BPF (CLE or LNA) with MIS 
in terms of mean squared error. We see a reduction in mean squared error of two orders 
of magnitude when comparing MIS with CH, across all experiments, and performance (across 
all metrics) of MIS with $N=500$ is comparable with the performance of CH when $N=10$. BPF-LNA 
generally outperforms BPF-CLE, although the difference is small. Running the BPF schemes 
generally requires twice as much computational effort than MIS, whereas CH is roughly 
three times slower than MIS. Even when this additional cost is taken into account, MIS 
cannot be recommended in this example.

Naturally, the performance of BPF will depend on the accuracy of the normal approximations 
used by the CLE and LNA. In particular, we expect these approximations to be unsatisfactory 
when species numbers are low. Moreover, when the conditioned jump process exhibits nonlinear 
dynamics, we expect the Euler approximation to be particularly poor. We therefore repeated 
the experiments of Table~\ref{tab:tabBD} with $N=500$, $x_{0}=10$, and $x_{t}$ as 
the lower 1\% quantile of $X_{t}|X_{0}=10$. Results are reported in Table~\ref{tab:tabBD2}. 
We see that in this case, MIS outperforms 
BPF and the performance of BPF-CLE worsens as $t$ increases, suggesting that a single step 
of the Euler approximation is unsatisfactory for $t>0.1$. Use of the conditioned hazard 
on the other hand appears fairly robust to different choices of $x_{0}$, $x_{t}$ and $t$. 
%Therefore, in the following examples, we compare the performance of MIS and CH when used 
%inside particle marginal Metropolis-Hastings (PMMH) schemes to perform parameter inference.  
         
\begin{table}[t]
	\centering
	\begin{tabular}{|l|c|c|c|}
	\hline	
Method	& $t=0.1$ & $t=0.5$ & $t=1$ \\
	\hline
MIS     &5000, 4747, 7.3$\times 10^{-5}$ &4998, 4426, 3.1$\times 10^{-5}$ &4999, 4500, 3.7$\times 10^{-5}$ \\
CH      &5000, 4979, 8.7$\times 10^{-6}$ &5000, 4963, 2.3$\times 10^{-6}$ &5000, 4965, 2.58$\times 10^{-6}$ \\
BPF-CLE &5000, 4131, 3.9$\times 10^{-4}$ &5000, 3013, 1.4$\times 10^{-3}$ &5000, 3478, 1.8$\times 10^{-3}$ \\
BPF-LNA &5000, 3946, 3.6$\times 10^{-4}$ &5000, 3667, 1.4$\times 10^{-4}$ &5000, 3639, 1.3$\times 10^{-4}$ \\
	\hline
	\end{tabular}
	\caption{$\sum_{i=1}^{m}I(\widehat{\pi}_{N}(x_{t}|x_{0})>0)$, $\textrm{ESS}(\widehat{\pi}^{1:m}_{N}(x_{t}|x_{0}))$ and 
$\textrm{MSE}(\widehat{\pi}^{1:m}_{N}(x_{t}|x_{0}))$, based on 5000 runs of MIS, CH, BPF-CLE and BPF-LNA. For MIS, 
the expected number of non-zero estimates (as obtained analytically) is reported. 
In all cases, $N=500$, $x_{0}=10$ and $x_{t}$ is the lower 1\% quantile of $X_{t}|X_{0}=10$.}\label{tab:tabBD2}
\end{table}

\subsection{Lotka-Volterra}
\label{sec:lvmodel}

We consider a simple model of predator and prey interaction comprising 
three reactions:
\[
\mathcal{R}_{1}:\, \mathcal{X}_{1} \longrightarrow 2\mathcal{X}_{1},\quad
\mathcal{R}_{2}:\, \mathcal{X}_{1}+\mathcal{X}_{2} \longrightarrow 2\mathcal{X}_{2},\quad
\mathcal{R}_{3}:\, \mathcal{X}_{2} \longrightarrow \emptyset.
\]
Denote the current state of the system by $X=(X_{1},X_{2})'$ where we have 
dropped dependence of the state on $t$ for notational simplicity. The 
stoichiometry matrix is given by
\[
S = \left(\begin{array}{rrr} 
1 & -1 & 0\\
0 & 1 & -1
\end{array}\right)
\]
and the associated hazard function is 
\[
h(X,c) = (c_{1}X_{1}, c_{2}X_{1}X_{2}, c_{3}X_{2})'.
\]

We consider three synthetic datasets consisting of 51 observations at 
integer times on prey and predator levels 
generated from the stochastic kinetic model using Gillespie's direct method and 
corrupted with zero mean Gaussian noise. The observation equation (\ref{obs_eq}) is 
therefore
\[
Y_{t} = X_{t}+\varepsilon_{t},
\]
where $X_{t}=(X_{1,t},X_{2,t})'$, 
$\varepsilon_{t}\sim\textrm{N}(0,\sigma^{2})$. We 
took $\sigma=10$ to construct the first dataset ($\mathcal{D}_{1}$), 
$\sigma=5$ to construct the second ($\mathcal{D}_{2}$) and $\sigma=1$ to give 
the third synthetic dataset ($\mathcal{D}_{3}$). In all cases we assumed $\sigma^{2}$ to be known. 
True values of the rate constants $(c_{1},c_{2},c_{3})'$ 
were taken to be 0.5, 0.0025, and 0.3 following \cite{BWK08}. 
We took the initial latent state as $x_{0}=(71,79)'$ assumed 
known for simplicity. Independent proper Uniform $U(-8,8)$ 
priors were ascribed to each $\log(c_i)$, denoted by $\theta_{i}$, 
$i=1,2,3$ and we let $\theta=(\theta_{1},\theta_{2},\theta_{3})'$ 
be the quantity for which inferences are to be made.

For brevity, we refer to the likelihood-free PMMH scheme (based on forward 
simulation only) as PMMH-LF, and the scheme based on the conditioned 
hazard proposal mechanism as PMMH-CH. As the ODEs governing the LNA solution 
are intractable, we focus on the CLE implementation of the bridge particle 
filter and refer to this scheme as PMMH-BPF. A pilot run of PMMH-LF was performed for each dataset 
to give an estimate of the posterior variance $\widehat{\textrm{Var}}(\theta)$, 
posterior median and 3 additional sampled $\theta$ values. We denote the 
variance of the noise in the log posterior by $\tau^{2}$ and chose the 
number of particles $N$ for each scheme so that $\tau^{2}\approx 2$ 
at the estimated posterior median and $\tau^{2}<4$ at the remaining sampled $\theta$ values 
(where possible). We updated $\theta$ using a Gaussian random walk with an innovation variance 
given by $\lambda \widehat{\textrm{Var}}(\theta)$, with the scaling parameter $\lambda$ 
optimised empirically, using minimum effective sample size (ESS$_{\textrm{min}}$) 
over each parameter chain. PMMH-BPF requires specification of a set of intermediate 
times at which resampling could be triggered. We found that resampling every 0.2 time 
units worked well. We also found that tempering the CLE approximation by raising 
each contribution $q(y_{t}|x_{t_{k}})$ to the power $\gamma$ performed 
better than using the CLE approximation directly (with $\gamma=1$). We took 
$\gamma=0.5, 0.2$ and $0.1$ for each dataset $\mathcal{D}_{1}$, $\mathcal{D}_{2}$ 
and $\mathcal{D}_{3}$ respectively. 
All schemes were run for $10^{5}$ iterations, except for PMMH-LF when using 
dataset $\mathcal{D}_{1}$, whose computational cost necessitated a shorter run of 50,000 
iterations. All algorithms are coded in C and run on a desktop computer with 
a 3.4GHz clock speed.

\begin{table}[t]
	\begin{center}
	\begin{tabular}{|l|cccccc|}
         \hline
         & $N$ & $\tau^{2}$ & Acc. rate& ESS($\theta_1,\theta_2,\theta_3$) & Time (s)  & ESS$_{\textrm{min}}/s$\\
	\hline	
        & \multicolumn{6}{|c|}{$\mathcal{D}_{1}$ ($\sigma=10$)} \\
 PMMH-LF  &230 &2.0 &0.15 &(3471, 3465, 3760) &17661 & 0.196 \\
 PMMH-CH  &50  &2.1 &0.14 &(3178, 3153, 3095) &18773 & 0.165 \\  
 PMMH-BPF &220  &2.0 &0.16 &(3215, 2994, 3121) &27874 &0.107 \\      
\hline
        & \multicolumn{6}{|c|}{$\mathcal{D}_{2}$ ($\sigma=5$)} \\
 PMMH-LF    &440 &2.0 &0.15 &(3482, 3845, 3784) &33808 &0.103 \\
 PMMH-CH &35  &2.0 &0.15 &(3581, 3210, 3204) &13341 &0.240 \\      
 PMMH-BPF &250  &1.9 &0.17 &(3779, 3887, 4110) &33436 &0.113 \\   
\hline
        & \multicolumn{6}{|c|}{$\mathcal{D}_{3}$ ($\sigma=1$)} \\
 PMMH-LF    &25000 &1.9 & 0.18 &(2503, 2746, 2472) & 1277834 &0.00193 \\
 PMMH-CH &55 &1.9 &0.14 &(2861, 2720, 2844) &22910 &0.118 \\      
 PMMH-BPF &3000 &1.8 &0.18 &(3732, 3990, 4221) &290000 &0.0129 \\ 
\hline
\end{tabular}
	\caption{Lotka-Volterra model. Number of particles $N$, variance of the noise in the log-posterior ($\tau^{2}$) at the posterior median, 
acceptance rate, effective sample size (ESS) of each parameter chain and wall clock time in seconds 
and minimum (over each parameter chain) ESS per second.}\label{tab:tabLV}
	\end{center}
\end{table}

\begin{figure}[t]
\centering
%\psfragscanon
\psfrag{thet1}[][][0.8][0]{$\theta_1$}
\psfrag{thet2}[][][0.8][0]{$\theta_2$}
\psfrag{thet3}[][][0.8][0]{$\theta_3$}
\includegraphics[width=5cm,height=15cm,angle=-90]{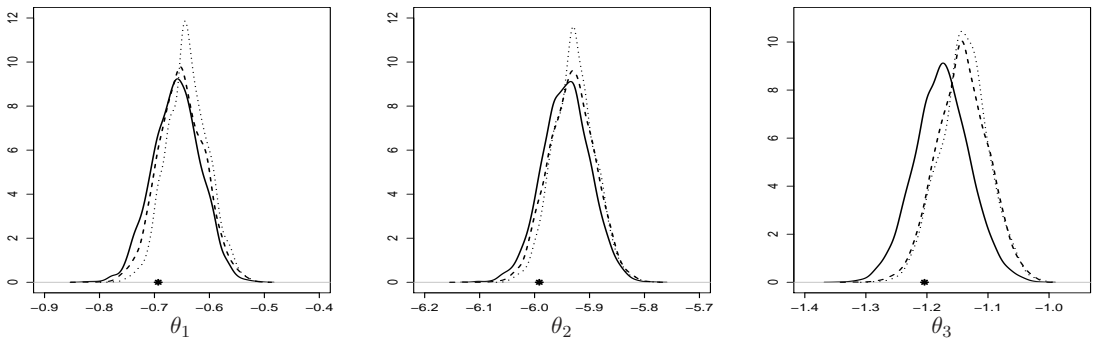}
\caption{Lotka-Volterra model. Marginal posterior distributions based on synthetic data generated using 
$\sigma^{2}=10$ (solid), $\sigma^{2}=5$ (dashed) and $\sigma^{2}=1$ (dotted). Values 
of each $\theta_{i}$ that produced the data are indicated.}
\label{fig:margpost}
\end{figure}

Figure~\ref{fig:margpost} shows the marginal posterior distributions for each dataset and Table~\ref{tab:tabLV} 
summarises the overall efficiency of each PMMH scheme. When using PMMH-CH, relatively few 
particles are required (ranging from 35--55) even as noise in the observation process reduces. Although 
PMMH-BPF required fewer particles than PMMH-LF, as $\sigma$ is reduced, increasing numbers of particles are required 
by both schemes to optimise overall efficiency. We measure overall efficiency by comparing 
minimum effective sample size scaled by wall clock time (ESS$_{\textrm{min}}/s$). When using 
$\mathcal{D}_{1}$ ($\sigma=10$), there is little difference in overall efficiency between each scheme 
although PMMH-LF is to be preferred. For dataset $\mathcal{D}_{2}$ ($\sigma=5$), PMMH-BPF and PMMH-LF 
give comparable performance whilst PMMH-CH outperforms PMMH-LF by a factors of 2.3. For 
$\mathcal{D}_{3}$ ($\sigma=1$) PMMH-CH and PMMH-BPF outperform PMMH-LF by factors of 61 and 6.7 
respectively. Computational cost precluded the use of PMMH-LF on a dataset with $\sigma<1$, 
however, our experiments suggest that PMMH-CH can be successfully applied to synthetic 
data with $\sigma=0.1$ by using just $N=50$ particles. Finally, we note that PMMH-CH appears 
to outperform PMMH-BPF, and, whereas the latter requires choosing appropriate 
intermediate resampling times and a tempering parameter $\gamma$, PMMH-CH requires 
minimal tuning. Therefore, in the following example, we focus on the PMMH-CH scheme.

\subsection{Motility regulation}
\label{sec:motmodel}
We consider here a simplified model of a key cellular decision made by 
the gram-positive bacterium \emph{Bacillus subtilis} \citep{sonenshein2002}. 
This decision is whether or not to grow flagella and become motile \citep{kearns2005}.
The \emph{B.\ subtilis}\ sigma factor $\sigma^D$ is key for the
regulation of motility. Many of the genes and operons encoding
motility-related proteins are governed by this $\sigma$ factor, and so
understanding its regulation is key to understanding the motility
decision. The gene for $\sigma^D$ is embedded in a large operon
containing several other motility-related genes, known as the
\emph{fla/che} operon. The \emph{fla/che} operon itself is under the
control of another $\sigma$ factor, $\sigma^A$, but is also regulated
by other proteins. In particular, transcription of the operon is
strongly repressed by the protein \emph{CodY}, which is encoded
upstream of \emph{fla/che}. \emph{CodY} inhibits transcription by
binding to the \emph{fla/che} promoter. Since \emph{CodY} is
upregulated in good nutrient conditions, this is thought to be a key
mechanism for motility regulation.
As previously mentioned, many motility-related genes are under the
control of $\sigma^D$. For simplicity we focus here on one such gene,
\emph{hag}, which encodes the protein \emph{flagellin} (or
\emph{Hag}), the key building block of the flagella. It so happens
that \emph{hag} is also directly repressed by \emph{CodY}. The
regulation structure can be encoded as follows.

\[
\begin{array}{ll}
\mathcal{R}_{1}:\, \codY \longrightarrow \codY+\CodY, & \quad\mathcal{R}_{2}:\, \CodY\longrightarrow \emptyset,\\  
\mathcal{R}_{3}:\, \flache\longrightarrow \flache+\SigD,&\quad \mathcal{R}_{4}:\, \SigD  \longrightarrow \emptyset,\\
\mathcal{R}_{5}:\, \SigDhag \longrightarrow \SigD+\hag+\Hag &\quad \mathcal{R}_{6}:\, \Hag \longrightarrow \emptyset,\\
\mathcal{R}_{7}:\, \SigD+\hag \longrightarrow \SigDhag, &\quad \mathcal{R}_{8}:\, \SigDhag\longrightarrow \SigD+\hag,\\  
\mathcal{R}_{9}:\, \CodY+\flache\longrightarrow \CodYflache,&\quad \mathcal{R}_{10}:\, \CodYflache  \longrightarrow \CodY+\flache,\\
\mathcal{R}_{11}:\, \CodY+\hag \longrightarrow \CodYhag &\quad \mathcal{R}_{12}:\, \CodYhag \longrightarrow \CodY+\hag. 
\end{array}
\]

\begin{figure}[t]
\centering
%\psfragscanon
\psfrag{t}[][][0.8][0]{$t$}
\psfrag{sigD}[][][0.8][0]{$\SigD$}
\psfrag{hag}[][][0.8][0]{$\hag$}
\psfrag{Hag}[][][0.8][0]{$\Hag$}
\psfrag{flache}[][][0.8][0]{$\flache$}
\psfrag{cody}[][][0.8][0]{$\codY$}
\psfrag{CodY}[][][0.8][0]{$\CodY$}
\psfrag{sigDhag}[][][0.8][0]{$\SigDhag$}
\psfrag{CodYhag}[][][0.8][0]{$\CodYhag$}
\psfrag{CodYflache}[][][0.8][0]{$\CodYflache$}
\includegraphics[width=4cm,height=15cm,angle=-90]{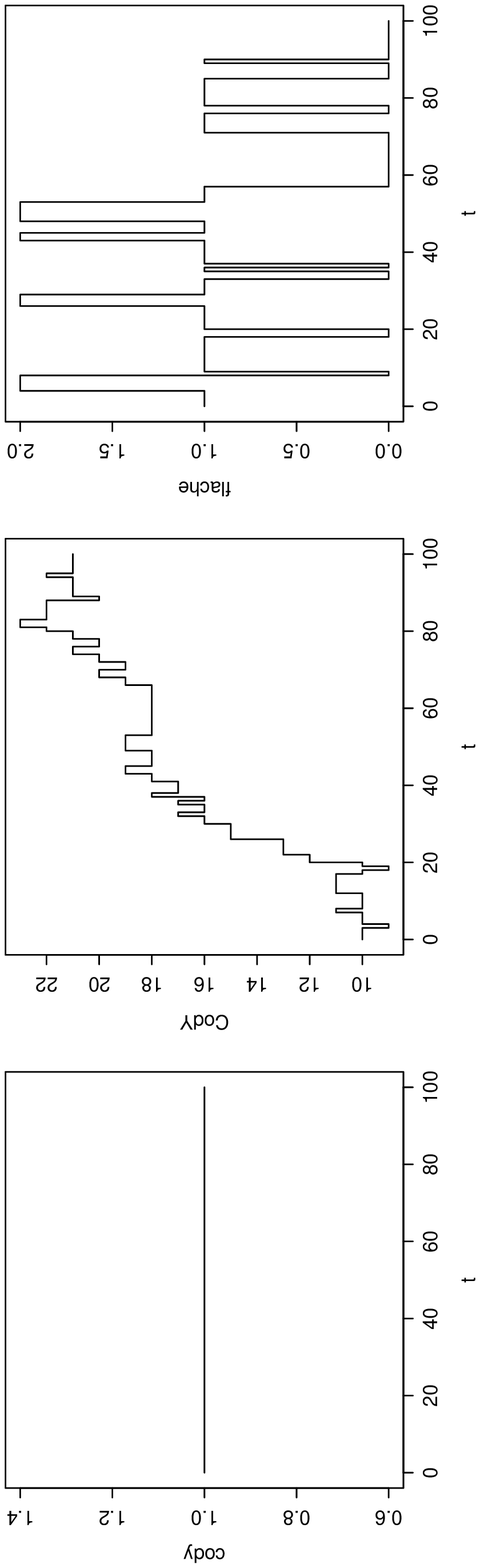}
\includegraphics[width=4cm,height=15cm,angle=-90]{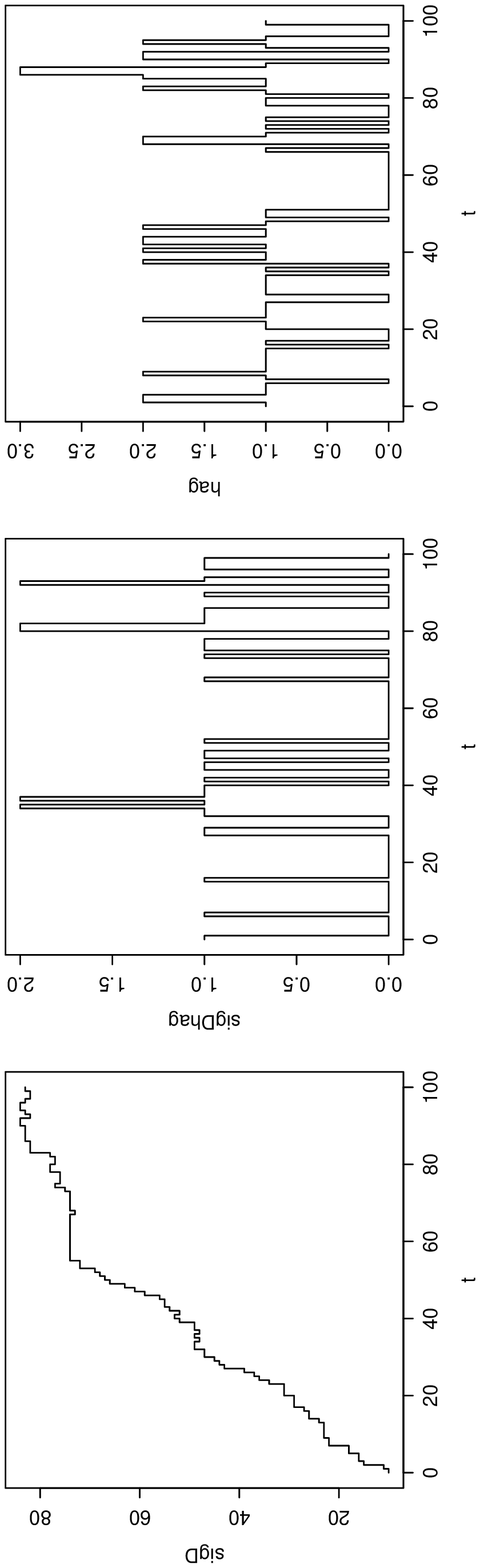}
\includegraphics[width=4cm,height=15cm,angle=-90]{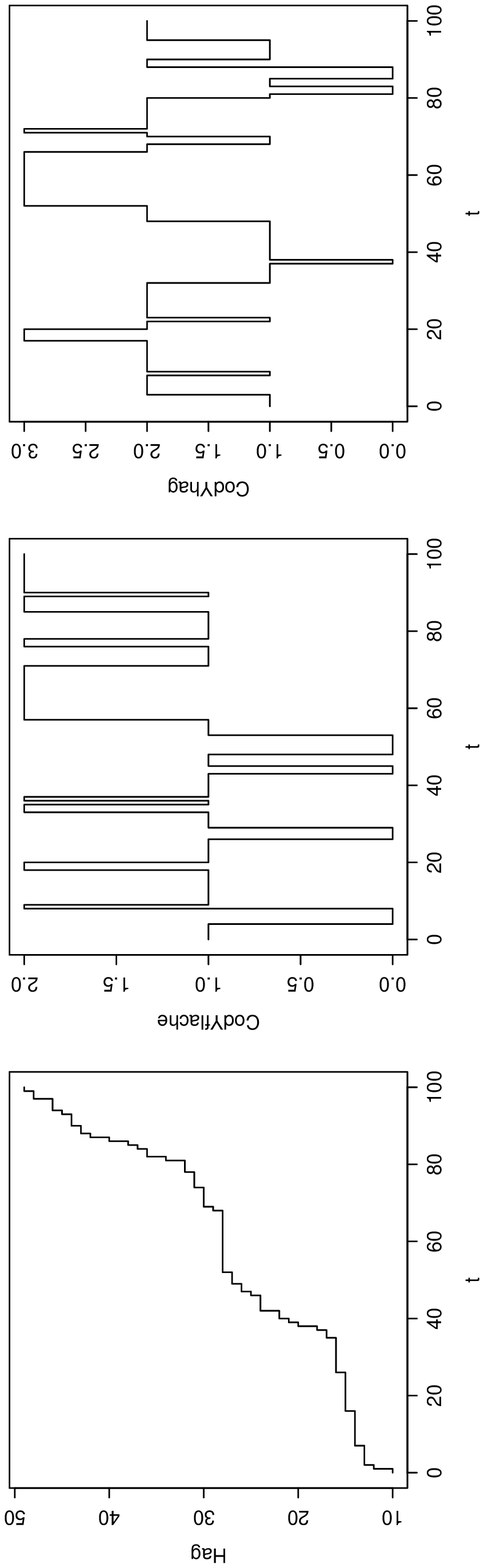}
\caption{A typical realisation of the motility model.}
\label{fig:motdat}
\end{figure}

Following \cite{Wilkinson11}, we assume that three rate constants are uncertain, namely $c_{3}$ 
(governing the rate of production of $\SigD$), $c_{9}$ and $c_{10}$ (governing the rate at which $\CodY$ 
binds or unbinds to the $\flache$ promoter). Values of the rate constants are taken to be 
\[
c=(0.1,0.0002,1,0.0002,1.0,0.0002,0.01,0.1,0.02,0.1,0.01,0.1)'
\] 
and initial values of $(\codY,\CodY,\flache,\SigD,\SigDhag,\hag,\Hag,\CodYflache,\CodYhag)$ are
\[
x_{0}=(1,10,1,10,1,1,10,1,1)'.
\]
Gillespie's direct method was used to simulate 3 synthetic datasets consisting of 
51 observations on $\SigD$ only, with inter-observation times of $\Delta t=1,2,5$ time units. A 
full realisation from the motility model that was used to construct each dataset is shown in 
Figure~\ref{fig:motdat}. The assumed initial conditions and parameter choices give inherently discrete 
time series.

To provide 
a challenging (but unrealistic) scenario for the PMMH-CH scheme we assume that error-free 
observations are available. We adopt independent proper Uniform priors on the log scale:
\begin{align*}
\log(c_{3})&\sim \textrm{U}\left(\log\{0.01\},\log\{100\}\right)\\
\log(c_{9})&\sim \textrm{U}\left(\log\{0.0002\},\log\{2\}\right)\\
\log(c_{10})&\sim \textrm{U}\left(\log\{0.001\},\log\{10\}\right)
\end{align*}
which cover two orders of magnitude either side of the ground truth. We ran PMMH-CH for 
$10^{5}$ iterations, after determining (from short pilot runs) suitable numbers of particles for 
each dataset, and a scaling $\lambda$ for use in the Gaussian random walk proposal kernel.

\begin{figure}[t]
\centering
%\psfragscanon
\psfrag{log(c3)}[][][0.8][0]{$\log(c_{3})$}
\psfrag{log(c9)}[][][0.8][0]{$\log(c_{9})$}
\psfrag{log(c10)}[][][0.8][0]{$\log(c_{10})$}
\includegraphics[width=5cm,height=15cm,angle=-90]{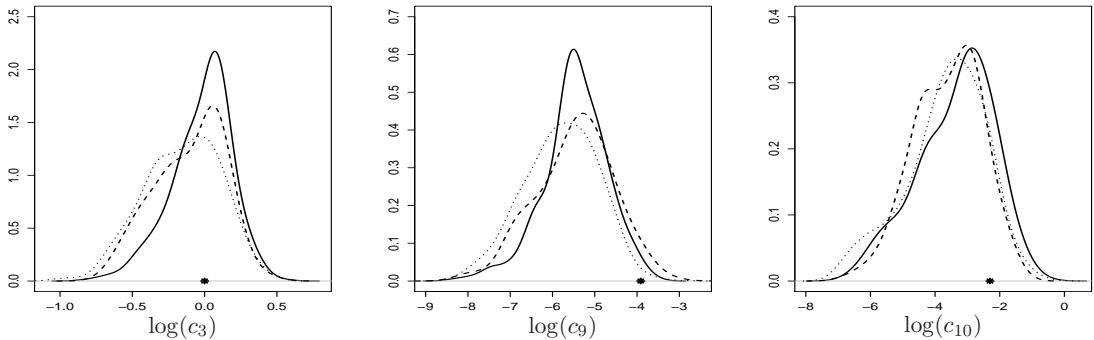}
\caption{Motility regulation model. Marginal posterior distributions based on synthetic data with 
inter-observation times of $\Delta t=1$ (solid), $\Delta t=2$ (dashed) and $\Delta t=5$ (dotted). Values 
of each $\log(c_{i})$ that produced the data are indicated.}
\label{fig:margpostMOT}
\end{figure}

\begin{table}[t]
	\begin{center}
	\begin{tabular}{|l|cccccc|}
         \hline
  &       $N$ & $\tau^{2}$ & Acc. rate& ESS($\theta_1,\theta_2,\theta_3$) & Time (s)  & ESS$_{\textrm{min}}/s$\\
	\hline	
        %\multicolumn{6}{|c|}{$\Delta t=1$} \\
$\Delta t=1$ & 400  &1.99 &0.10 &(1635, 2156, 1625) &6933 &0.23 \\
%\hline
 %\multicolumn{6}{|c|}{$\Delta t=2$} \\
$\Delta t=2$ & 600  &2.05 &0.10 &(1870, 1215, 1518) &6950 &0.17 \\
%\hline
 %\multicolumn{6}{|c|}{$\Delta t=5$} \\
$\Delta t=5$ & 1200  &2.01 &0.06 &(797, 791, 673) &13628 &0.05 \\
\hline
\end{tabular}
	\caption{Motility regulation model. Number of particles $N$, variance of the noise in the log-posterior ($\tau^{2}$) at the posterior median, 
acceptance rate, effective sample size (ESS) of each parameter chain and wall clock time in seconds 
and minimum (over each parameter chain) ESS per second.}\label{tab:tabMR}
	\end{center}
\end{table}

Figure~\ref{fig:margpostMOT} shows the marginal posterior distributions for each dataset. 
We see that despite observing levels of $\SigD$ only, sampled parameter values 
are consistent with the ground truth. Table~\ref{tab:tabMR} summarises the overall efficiency of PMMH-CH when 
applied to each dataset. We see that as the inter-observation time $\Delta t$ increases, 
larger numbers of particles are required to maintain a variance in log-posterior 
of around 2 at the estimated posterior median. Despite using increased particle 
numbers, statistical efficiency, as measured by effective sample size, appears to 
reduce as $\Delta t$ is increased. We observed that parameter chains were 
more likely to ``stick'' (and note the decreasing acceptance rate) leading to 
reduced ESS. This is not surprising given the assumptions used to derive the conditioned hazard, and 
we expect its performance to diminish as inter-observation time increases.     

\section{Discussion and conclusions}
\label{sec:disc}

This paper considered the problem of performing inference for the 
parameters governing Markov jump processes in the presence of 
informative observations. Whilst it is possible to construct 
particle MCMC schemes for such models given time course data that 
may be incomplete and subject to error, the simplest ``likelihood-free'' 
implementation is likely to be computationally intractable, except in high 
measurement error scenarios. To circumvent this issue, we have proposed 
a novel method for simulating from a conditioned jump process, 
by approximating the expected number of reactions between observation 
times to give a conditioned hazard. We find that a simple implementation 
of this approach, with exponential waiting times between proposed 
reaction events, works extremely well in a number of scenarios, and even in challenging 
multivariate settings. It should be noted however, that the assumptions under-pinning 
the construct are likely to be invalidated as inter-observation time increases. 
We compared this approach with a bridge 
particle filter adapted from \cite{delmoral14}. Implementation of this 
approach requires the ability to simulate from the model and access to an 
approximation of the (unavailable) transition probabilities. The overall 
efficiency of the scheme depends on the accuracy and computational cost 
of the approximation. Use of the LNA inside the bridge particle filter 
appears promising, although the requirement of solving a system of ODEs 
for each particle, and whose dimension increases quadratically with the 
number of species, is likely to be a barrier to its successful application 
in high dimensional systems. Using a numerical approximation to the CLE 
offers a cheaper but less accurate alternative. The bridge particle filter 
(based on either the LNA or CLE) requires specification of appropriate 
intermediate resampling times and, when the approximations are likely 
to be light tailed relative to the jump process transition probability, 
a tempering parameter. Use of the conditioned hazard on the other hand 
requires minimal tuning. This approach was successfully applied to the 
problem of inferring the rate constants governing a Lotka-Volterra 
system and a simple model of motility regulation.   

Improvements to the proposed algorithms remain of interest and are the subject of 
ongoing research. For example, when using the bridge particle filter, it may be 
possible to specify a resampling regime dynamically, based on the expected 
time to the next reaction event, evaluated at, for example, the LNA mean. 
An exact implementation of the conditioned hazard approach and the potential 
improvement it may offer is also of interest, especially for systems with 
finite state space, which would permit a thinning approach \citep{lewis1979} to reaction 
event simulation.

%\section*{Acknowledgements}

\bibliographystyle{apalike}
\bibliography{bridgebib}

\appendix

%\section{Appendices}

\end{document}